\documentclass{PoS}
\newcommand{\tm}[1]{\mathrm{ #1}}

\title{Applications of Holography to Strongly Coupled Plasmas}

\ShortTitle{Applications of Holography to Strongly Coupled Plasmas}

\author{Carlo Ewerz, \speaker{Konrad Schade}%
  \\
  Institut f\"ur Theoretische Physik, Universit\"at
  Heidelberg,\\Philosophenweg 16, 69120 Heidelberg, Germany\\%
  \textup{and}\\%
  ExtreMe Matter Institute EMMI, GSI Helmholtzzentrum f\"ur
  Schwerionenforschung,\\Planckstra\ss{}e 1, 64291 Darmstadt,
  Germany\\%
  E-mail:
  \begin{minipage}[t]{.8\linewidth}
    \email{c.ewerz@thphys.uni-heidelberg.de},\\%
    \email{k.schade@thphys.uni-heidelberg.de}
  \end{minipage}}

\abstract{
We study several observables related to heavy quarks in 
strongly coupled plasmas using the gauge/gravity correspondence. 
Besides the AdS${}_5$ space dual to $\mathcal{N}\!=\!4$ supersymmetric 
Yang-Mills theory we consider large classes of theories obtained from 
various deformations of the AdS${}_5$ space. Among them are theories that 
solve equations of motion of a 5-dimensional Einstein-Hilbert-scalar 
action. 
Specifically, we calculate the screening distance of a heavy 
quark-antiquark pair moving at constant velocity through the plasma, 
the running coupling defined via the free energy of such a static pair, 
and the energy radiation from a heavy quark forced into a circular 
motion in the plasma. We find that these observables show universal 
behaviour in large classes of theories. The screening distance in 
these classes of theories, that is 
the maximal distance for which a heavy quark-antiquark pair is bound, 
is found to be bounded from below by its value in $\mathcal{N}\!=\!4$ 
super Yang-Mills theory. 
}

\FullConference{Xth Quark Confinement and the Hadron Spectrum,\\
		October 8-12, 2012\\
		TUM Campus Garching, Munich, Germany}

\begin{document}

\section{Introduction}
\label{sec:intro}

Since its discovery the gauge/gravity (or AdS/CFT) correspondence 
\cite{Maldacena:1997re,Gubser:1998bc,Witten:1998qj}
has been used to explore various aspects of 
supersymmetric gauge theories at strong coupling, including their 
behavior at finite temperature. More recently, it was 
realized that this correspondence or duality has the potential to help answer  
questions also in the realm of various strongly coupled systems that 
are accessible to experiment. It turned out that a particularly interesting 
and fruitful application of the AdS/CFT correspondence is to 
the quark-gluon plasma created and studied at RHIC and LHC. 
These experiments find strong evidence that the quark-gluon 
plasma is in fact strongly coupled. New theoretical methods like 
the AdS/CFT correspondence are especially welcome here 
since the strong coupling makes ab initio 
calculations based on QCD very hard, leaving only lattice QCD 
as a reliable tool. Lattice QCD, however, can only address static 
properties reliably while it has difficulties in the calculation of 
many dynamical problems that are of interest in the study of 
the quark-gluon plasma. The AdS/CFT correspondence can address 
not only dynamical as well as static observables but in 
addition has the advantage that its results are obtained 
from computationally simple calculations on the gravity side. 

In its original form, the AdS/CFT correspondence relates 
4-dimensional $\mathcal{N}\!=\!4$ supersymmetric Yang-Mills 
theory (SYM) with gauge group $\mbox{SU}(N_c)$ 
at strong coupling and taken in the large-$N_c$ limit to 
weakly coupled (super)gravity on a 5-dimensional Anti-de Sitter 
space AdS${}_5$. The 4-dimensional theory can be thought of 
as living `on the boundary' of the AdS space, so that the 
equivalence of these two descriptions of the same 
underlying physics can actually be understood as a realization 
of the holographic principle. 
Finite temperature on the gauge theory side is implemented in 
the correspondence by putting a black hole in the bulk of the 
AdS${}_5$ space, such that the temperature of the gauge theory coincides 
with the Hawking temperature of the black hole. 

But QCD and $\mathcal{N}\!=\!4$ SYM are two very different theories. A successful 
application of the correspondence to real-world QCD therefore 
appears far from obvious. 
At a finite temperature above the deconfinement temperature $T_c$ 
the situation is already more promising. On the 
one hand, chiral symmetry and confinement are lost in QCD. On the 
other hand, the supersymmetry and the conformal symmetry 
of $\mathcal{N}\!=\!4$ SYM are broken by finite temperature. 
One can therefore hope that QCD and SYM share at least some 
properties at finite temperature. 

Considerable efforts have been made to find a holographic description 
of theories that are even closer to QCD than $\mathcal{N}\!=\!4$ SYM. 
This can be achieved by considering deformations of the AdS space 
which can be introduced in various ways. A common feature of these 
deformations is that they in some way or another break the conformal 
invariance of $\mathcal{N}\!=\!4$ SYM explicitly. 
One line of thought is to 
construct theories that reproduce as many properties of QCD as possible, 
although an exact holographic dual of QCD would certainly be extremely 
difficult to find, and in fact not even known to exist. In a second 
line of thought -- the one which we will follow here -- one 
considers large classes of deformations and 
tries to identify universal properties of observables in these 
classes of holographic theories. An example of this approach is 
the well-known finding that the ratio of shear viscosity to entropy 
density $\eta/s$ equals $1/(4\pi)$ in large classes of theories 
\cite{Policastro:2001yc,Kovtun:2004de}. Other observables might 
change only very little or might consistently change in a certain 
direction under the influence of deformations of the AdS space. 
If such a universal property is found one can hope that the 
behavior of QCD is not too different from the trend observed 
in these large classes of theories. 
In this talk we would like to present some interesting results obtained 
in this approach. A more detailed description is given elsewhere 
\cite{Schade:2012mqa}. 

\section{AdS-type Gravity Models for Strongly Coupled Plasmas at Finite Temperature}
\label{sec:models}

The gravity dual of $\mathcal{N}\!=\!4$ SYM at finite temperature is 
AdS${}_5$ with a black hole at a position $z_h$ in the 5th dimension, 
\begin{equation}
\label{n4metric}
\tm{d}s^2 = \frac{L_{\rm AdS}^2}{z^2} \, \left[ - f(z) \, \tm{d}t^2 
+ \tm{d} \vec{x}^2 + \frac{\tm{d}z^2}{f(z)} \right] 
\end{equation}
with 
\begin{equation}
\label{n4h}
f(z)= 1 - \frac{z^4}{z_h^4} \,.
\end{equation}
The AdS curvature radius $L_{\rm AdS}$ is assumed to be large, corresponding 
to a large 't Hooft coupling of the gauge theory. The temperature of the 
dual $\mathcal{N}\!=\!4$ gauge theory is $T=1/(\pi z_h)$. 

In order to construct theories closer to real-world QCD one can consider 
various modifications of this metric. We will consider classes of theories 
with metrics of the general form 
\begin{equation}
\label{eq.ObsPhysQuant:RunCoup:Screen:Metric}
\tm{d} s^2 = \tm{e}^{2 A(z)}  \left( - h(z) \tm{d} t^2 + \tm{d} \vec{x}^{\, 2} \right) 
+ \tm{e}^{2 B(z)} \, \frac{\tm{d} z^2}{h(z)} \,. 
\end{equation}
The corresponding temperature of the dual 4-dimensional theory is 
\begin{equation}
\label{eq.NonconfMeMo:SWT:Temp:Tgen}
T = e^{A(z_h) - B(z_h)} \, \frac{| h'(z_h)|}{4 \pi} \,,
\end{equation}
where $z_h$ is again the position of the black hole horizon in the bulk 
of AdS${}_5$, $h(z_h)=0$. 
The $\mathcal{N}\!=\!4$ case (\ref{n4metric}) is recovered for $A(z)=B(z)=0$ 
and $h(z)=f(z)$. 

A simple class of deformations of the $\mathcal{N}\!=\!4$ case was 
proposed in \cite{Andreev:2006ct} and \cite{Kajantie:2006hv} motivated 
by the widely discussed soft-wall model at zero temperature 
\cite{Karch:2006pv}. This one-parameter class of metrics 
is obtained by either starting with a $T=0$ soft-wall model and 
inserting the black-hole factor (\ref{n4h}) in the same way as 
in (\ref{n4metric}). Alternatively, one can with the same result start from 
(\ref{n4metric}) and multiply it by a soft-wall factor. We therefore 
call this the finite temperature soft wall model, SW${}_T$. 
Its metric (in the string frame, see below) has the form 
\begin{equation}
\label{eq.Linear:ExplLinear:SWTmetricStringframe}
\tm{d} s^2 =  \frac{L_{\rm AdS}^2}{z^2} \, \tm{e}^{\frac{29}{20} c \, z^2} \, 
\left( - f(z) \,  \tm{d}t^2 + \tm{d}\vec{x}^{\,2} + \frac{\tm{d}z^2}{f(z)} \right)\,.
\end{equation}
The dimensionful 
parameter $c$ determines the non-conformality of the SW${}_T$ model. 
According to 
(\ref{eq.NonconfMeMo:SWT:Temp:Tgen}) the temperature of the SW${}_T$ 
model is again $T=1/(\pi z_h)$. The main drawback of the SW${}_T$ model 
is that it is not a solution of a 5-dimensional gravitational action. 
Therefore its consistency as a theory is somewhat doubtful. 

A class of consistent deformations of the AdS${}_5$ metric (\ref{n4metric}) 
can be constructed as solutions of a 5-dimensional Einstein-Hilbert-scalar 
action. We follow \cite{Gubser:2008ny,DeWolfe:2009vs} and start 
with the action 
\begin{equation}
\label{eq.NonconfMeMo:ConDefMeMo:5DEHdaction}
S_\tm{EHs} = \frac{1}{16 \pi G^{(5)}_\tm{N}} \, \int \tm{d}^5 x \, 
\sqrt{- G} \, \left( \mathcal{R} - \frac{1}{2} ( \partial \Phi )^2 
- V(\Phi) \right) \,,
\end{equation}
where $G_N^{(5)}$ is the 5-dimensional Newton constant, 
$\mathcal{R}$ is the Ricci scalar, and $V(\Phi)$ a potential for 
the scalar field $\Phi(z)$. $\Phi(z)$ can but need not be the dilaton. 
We consider both possibilities as independent models which have 
been termed `string frame' and `Einstein frame' model, respectively 
\cite{DeWolfe:2009vs}. 
In order to obtain metrics that resemble the soft-wall type models 
one can consider classes of 2-parameter models of the form 
(\ref{eq.ObsPhysQuant:RunCoup:Screen:Metric}) by making the 
following ansatz. We switch to a gauge where the bulk coordinate is 
identified with the scalar $\Phi$ such that our general metric is 
\begin{equation}
\label{eq.NonconfMeMo:ConDefMeMo:1Para:2ParaMetricAnsatz}
\tm{d} s^2 = \tm{e}^{2 A(\Phi)} \, \big(- h(\Phi) \tm{d} t^2 
+ \tm{d} \vec{x}^{\,2} \big) + \tm{e}^{2 B(\Phi)} \, \frac{\tm{d} \Phi^2}{h(\Phi)} \, .
\end{equation}
We then choose $A(\Phi)$ and the scalar $\Phi(z)$ as 
\begin{equation}
\label{2pA}
 A(\Phi) = \frac{1}{2} \, \log \left( \frac{L_{\rm AdS}^2}{z^2} \right) - \frac{1}{2} c z^2 \,,
\quad\quad\quad\quad\quad
 \Phi = \sqrt{\frac{3}{2}} \phi z^2 
\end{equation}
and determine $B(z)$, the horizon function $h(z)$, and the potential 
$V(\Phi)$ from the equations of motion 
of the action (\ref{eq.NonconfMeMo:ConDefMeMo:5DEHdaction}). The 
resulting class of metrics has $c$ and $\alpha \equiv c/\phi$ as two 
independent parameters characterizing the deformation 
\cite{DeWolfe:2009vs}. The temperature 
is given by a formula analogous to (\ref{eq.NonconfMeMo:SWT:Temp:Tgen}) 
involving the horizon position. Note that we do not fix the scalar potential 
$V(\Phi)$ from the beginning but use the freedom to adjust it such that 
the solution has the desired form (\ref{2pA}). The class of 2-parameter 
metrics obtained in this way defines consistent 4-dimensional theories 
via the holographic duality, but we should point out that in general we 
do not know their Lagrangian. 

\section{Screening Distance and Running Coupling}
\label{sec:screening}

We first consider the screening of a heavy quark-antiquark pair 
moving with velocity $v$ in a plasma. This problem was first addressed for 
$\mathcal{N}\!=\!4$ SYM in \cite{Liu:2006nn} and for 
the SW${}_T$ model in \cite{Liu:2008tz}. 
The free energy $F(L)$ of such a system is given 
by the thermal expectation value of a timelike Wegner-Wilson loop, 
$\langle W(\mathcal{C}) \rangle = \exp \left[ - i \mathcal{T} \, F(L) \right]$ 
where $L$ is separation between the quark and antiquark, and $\mathcal{T}$ 
the large temporal extension of the loop. In the gravity dual, this  
situation is described by an open string connecting the quark and antiquark 
and hanging down into the bulk. For infinitely heavy quarks the ends of 
the string are at $z=0$. The expectation value can then be obtained as 
\begin{equation}
\label{Wexpectgrav}
\langle W(\mathcal{C}) \rangle \propto e^{ i (S - S_0)} \,, 
\end{equation}
where $S$ is the Nambu-Goto action for the string worldsheet parametrized 
by $\sigma$ and $\tau$, 
\begin{equation}
\label{SNG}
S_\tm{NG} = - \frac{ 1}{2 \pi \alpha'} \, \int \tm{d} \sigma \tm{d} \tau  \, \sqrt{- \det g_{a b}} \, ,
\end{equation}
Here $g_{a b} = G_{\alpha \beta} \, \partial_a X^\alpha \, \partial_b X^\beta$ is 
the induced metric and $\alpha'$ is the string slope. Further, $S_0$ is 
twice the action of a single quark, given by the Nambu-Goto action of 
an open string hanging down into the bulk. The calculation of the 
actions $S$ and $S_0$ has to be performed in the string 
frame. For the 2-parameter model that means that the Einstein frame 
metric discussed above is multiplied by an exponential of the dilaton, 
$G^s_{\alpha\beta}= \exp(-\sqrt{2/3}\Phi) G^E_{\alpha\beta}$ in our 
normalization of the dilaton. If the scalar $\Phi$ in the action 
(\ref{eq.NonconfMeMo:ConDefMeMo:5DEHdaction}) is not a dilaton, 
the string frame and the Einstein frame metrics coincide. 

For a given distance $L$ between the quark and antiquark up to a 
maximally possible distance $L_{\rm max}$ there are always two 
string solutions. They can be parametrized by the maximal depth $z_c$ 
that they reach in the bulk. 
For the simple case that the pair is not moving with respect to 
the plasma, the relation between $L$ and $z_c$ is given by 
\begin{equation}
\label{eq.ObsPhysQuant:RunCoup:qqbarDist}
 L \pi T = 
 2 \pi T \int_0^{z_c}  \sqrt{ \frac{h(z_c) \, 
e^{2 B(z) + 4 A(z_c)-2A(z)}}{ h(z)^2 \, e^{4 A(z)} - h(z) \, h(z_c) \, e^{4 A(z_c)}}} \, 
\tm{d}z\, ,
\end{equation}
and a more complicated relation applies for $v\neq 0$. 
The solution with the larger $z_c$, that is the one coming closer to the 
horizon, turns out to have the higher energy and is hence the unstable 
solution. At distances larger than $L_{\rm max}$ there are no real-valued 
string solutions connecting the quark and antiquark. We call this 
maximally possible distance the screening distance. It should not be 
confused with the Debye screening length which characterizes the 
exponential flattening of the free energy at distances much larger than 
$L_{\rm max}$ where the quark and antiquark are no longer bound. 
(Also the Debye screening length can be addressed in the holographic 
setup, see \cite{Bak:2007fk}.) 

\begin{figure}[ht]
  \centering
  \includegraphics[width=.75\linewidth]{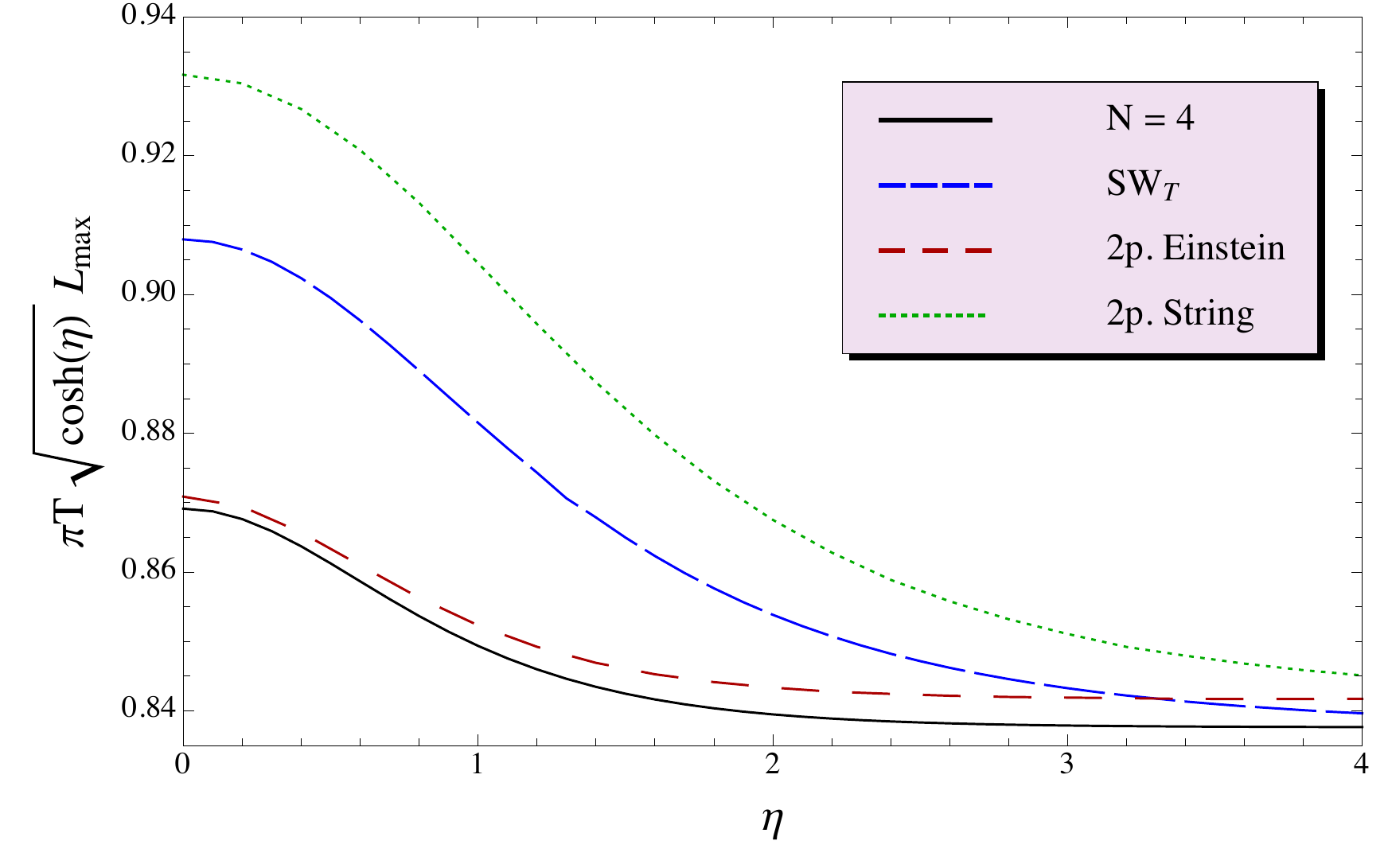}
  \caption{Dimensionless screening distance $\pi T L_{\rm max} \sqrt{\cosh \eta}$ 
of the $Q\bar{Q}$ pair in the plasma as a function of rapidity $\eta$. 
  \label{fig1}}
\end{figure}

We have calculated the screening distance $L_{\rm max}$ in all models 
discussed above for all its possible kinematical parameters, namely 
the velocity $v=\tanh \eta$ of the pair in the plasma and its 
orientation angle $\theta$ with respect to the velocity. Figure \ref{fig1} 
shows the dependence of $L_{\rm max}$ on the rapidity $\eta$ 
for the case $\theta=0$ for various models with typical values for the 
deformation parameters.  
We have chosen to plot the dimensionless quantity $\pi T L_{\rm max}$ 
for convenience and have divided out the dominant factor $1/\sqrt{\cosh \eta}$ 
found in the $\eta$-dependence. 

We have made the interesting observation that in all deformations in the 
above classes of models and for all kinematical parameters the screening 
distance $L_{\rm max}$ turns out to be larger than the corresponding value in 
$\mathcal{N}\!=\!4$ SYM. This leads us to conjecture that the screening 
distance in $\mathcal{N}\!=\!4$ SYM is a lower bound for all theories 
described by gravity duals. A general proof of this conjecture 
appears difficult. However, we have performed an analytic study of 
the problem for small perturbations of the metric around pure AdS${}_5$ 
space. It indicates that the conjecture holds in first order in the 
deformation of the metric under very mild and general assumptions 
\cite{Schade:2012mqa}. 

\begin{figure}[ht]
  \centering
  \includegraphics[width=.75\linewidth]{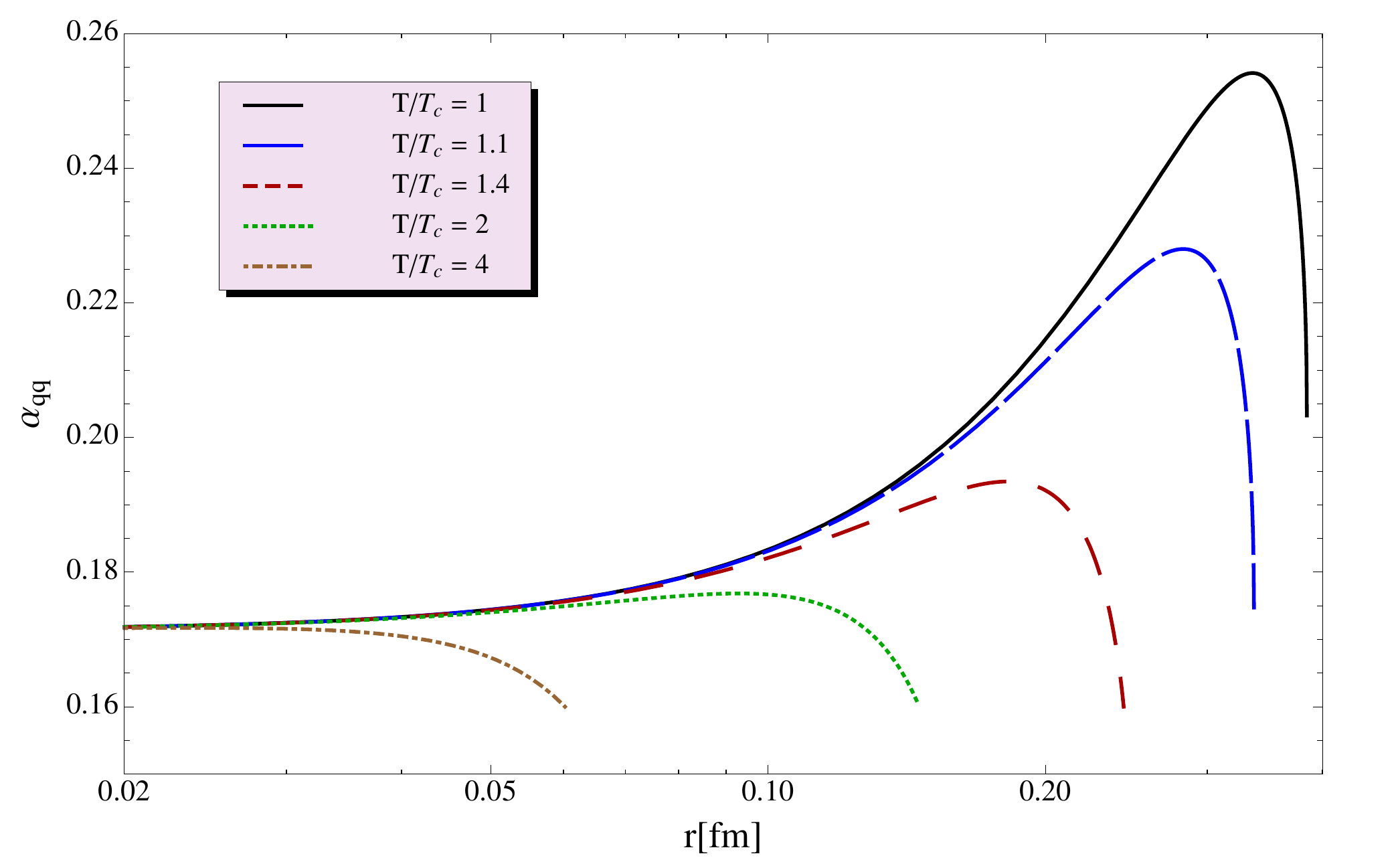}
  \caption{Running coupling $\alpha_{qq}$ versus the $Q\bar{Q}$-distance for 
several values of the temperature $T/T_c$, $T/T_c = 1, 1.1, 1.4, 2$ and $4$ 
and for fixed deformation $c/T^2$, $c/T^2 = 4$ in the SW${}_T$ model.
  \label{fig2}}
\end{figure}

The free energy $F(L)$ of the quark-antiquark pair can be extracted from 
(\ref{Wexpectgrav}) by computing $S$ and $S_0$. $S_0$ has been 
calculated in $\mathcal{N}\!=\!4$ SYM in \cite{Herzog:2006gh,Gubser:2006bz} 
and for the SW${}_T$ model in \cite{Nakano:2006js}. The extension to 
the models discussed here is straightforward. Once the free energy 
is known, one can proceed and calculate the running coupling in the whole 
class of models. It can be defined in terms of the free energy as 
\begin{equation}
\label{eq.ObsPhysQuant:RunCoup:CouplingDef}
\alpha_{qq} (L, \, T) \equiv \frac{3}{4} \, L^2 \, \frac{\tm{d} F(L, \, T)}{\tm{d} L} \, ,
\end{equation} 
where the normalization is motivated by the Casimir factor of QCD. 
In our setup the Lagrangian and hence the Casimir factor of the dual theory 
is not known, however. The normalization of the running coupling is further 
related to the 't Hooft coupling of the theory. As a consequence we treat 
the normalization of the result in our models as a free parameter to be 
suitably fixed. 
In figure \ref{fig2} we show as an example the running coupling in the 
SW${}_T$ model for various temperatures. 
We have chosen $T_c= 176 \,\mbox{MeV}$ for definiteness. 
The endpoint of the curves is always given by the screening distance 
$L_{\rm max}$. It is a universal property of the running coupling $\alpha_{qq}$ 
in all non-conformal models that it develops a maximum before falling rather 
steeply close to $L_{\rm max}$. We further find that in the 2-parameter model 
it is possible to adjust the model parameters such that the running 
coupling quite closely resembles the lattice data for QCD found in 
\cite{Kaczmarek:2004gv}. 

\section{Energy Loss of a Rotating Quark}
\label{sec:rotquark}

The energy loss of an accelerated quark in a medium can give valuable 
information about the properties of the medium itself. We therefore 
consider the motion of a quark on a circle of radius $R_0$ at 
constant angular velocity $\omega$. This is certainly not a realistic 
situation in any experiment, but serves well as the simplest model 
for an accelerated motion to be treated in the gauge/gravity 
correspondence. The problem had been addressed so far only in 
$\mathcal{N}\!=\!4$ SYM: at zero temperature in \cite{Athanasiou:2010pv}, 
and at finite temperature in \cite{Fadafan:2008bq}. We have calculated 
the motion of the string attached to the rotating quark also in the 
deformed metrics discussed above \cite{Schade:2012mqa}. 

\begin{figure}[ht]
  \centering
  \includegraphics[width=.48\linewidth]{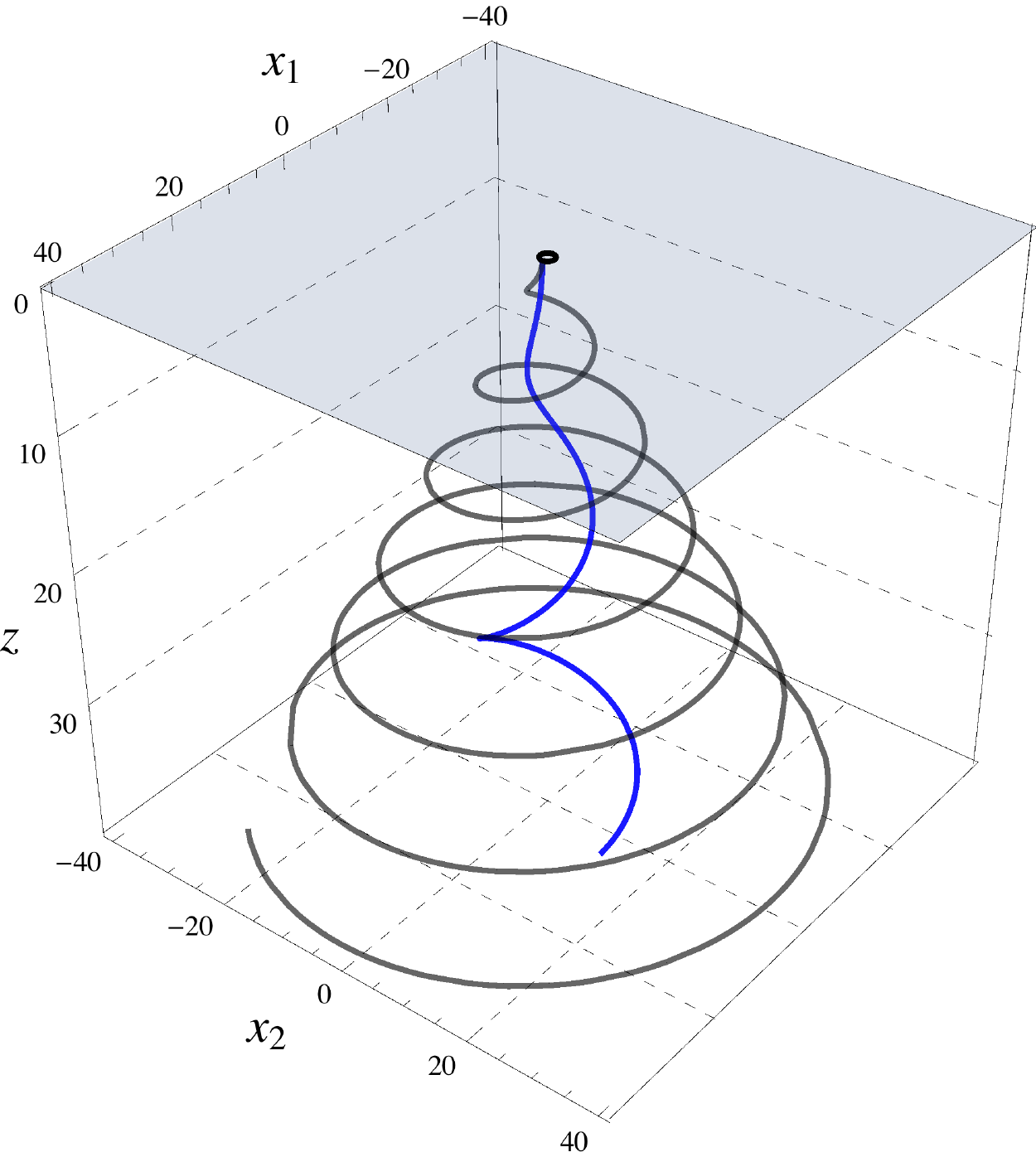}
  \hfill
  \includegraphics[width=.48\linewidth]{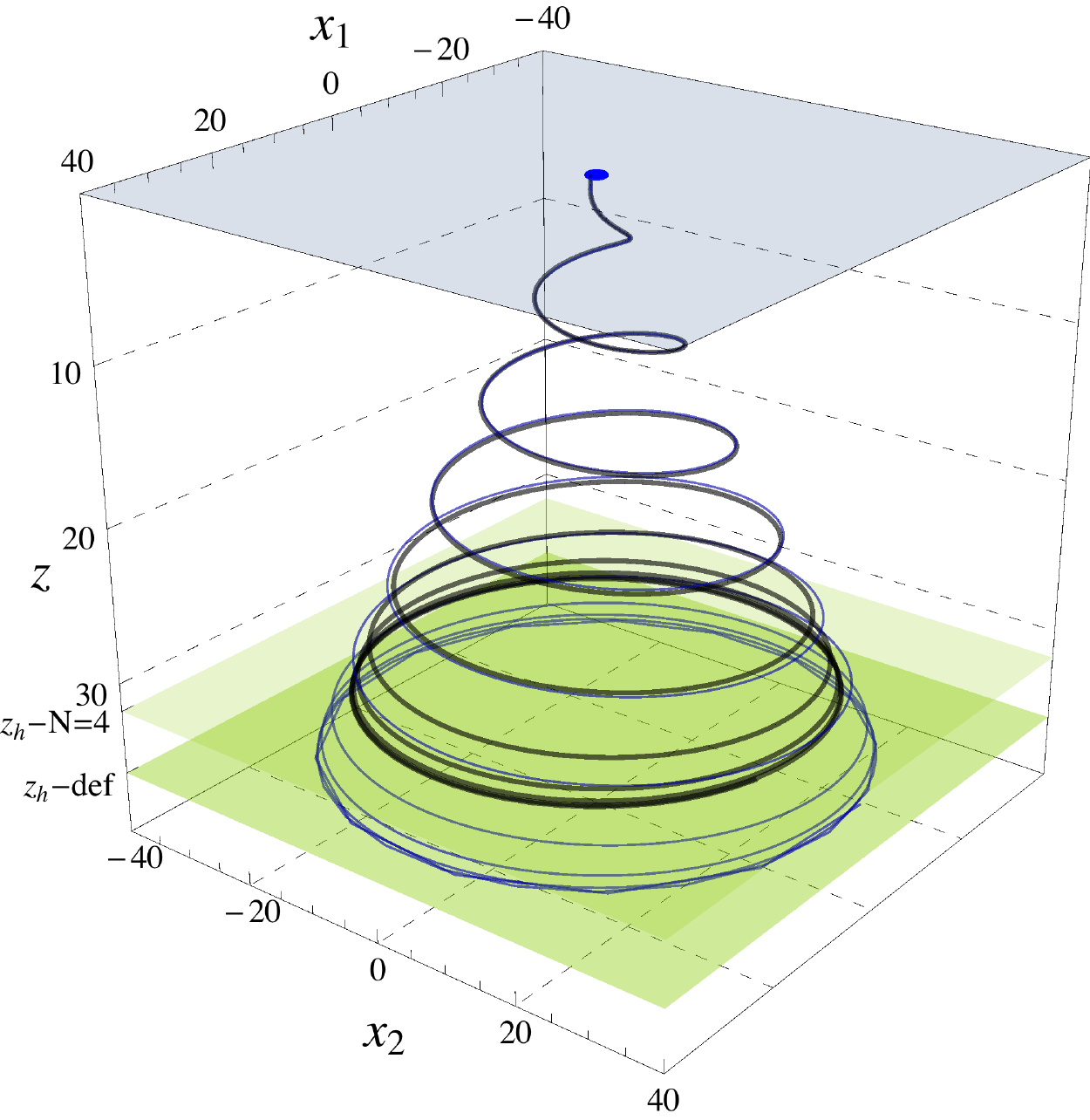}
  \caption{Left: String configuration in pure AdS${}_5$ 
($\mathcal{N}\!=\!4$ SYM at zero temperature) and 
radius $R_0 = 1$ and angular velocity $\omega = 0.3$ (blue) and $0.7$ (black). 
The small black circle at the top represents the trajectory of the rotating quark. 
Right: String configuration in 2-parameter string-frame model (blue line) for 
typical values of $\alpha$ and $c$ together with the corresponding 
$\mathcal{N}\!=\!4$ case (black line) at $T=0.01$, $R_0 = 1$ and $\omega = 0.7$ 
(in suitable dimensionless units). 
  \label{fig3}}
\end{figure}

In figure \ref{fig3} some typical configurations are shown. The left panel 
shows the pure AdS${}_5$ case at $T=0$ for two different angular 
velocities. In the $T=0$ case there is no horizon and the string 
spirals out to infinite radii deep in the bulk. 
This situation changes at finite temperature due to the horizon. 
Here the string extends only to a finite radius, see right panel. 

For the deformed models we in general observe that the string solution 
deviates considerably from the one corresponding to $\mathcal{N}\!=\!4$ SYM 
only at larger depths in the bulk. A more detailed analysis further shows that 
the energy radiated from the quark is mainly determined by the uppermost 
part of string configuration. As a consequence one finds that the radiation 
pattern in the deformed theories is in general very close to the one 
in $\mathcal{N}\!=\!4$ SYM. 

\section{Summary}
\label{sec:summary}

We have used holographic methods to study various observables 
related to the motion of heavy quarks in the plasma of a variety of theories at 
finite temperature. We have in particular considered the screening 
distance, that is the maximal distance for which a heavy quark and 
antiquark form a bound state, for all velocities and orientation 
angles of the pair in the plasma. The value of the screening distance 
in $\mathcal{N}\!=\!4$ SYM is a lower bound for a large class of 
theories for all kinematical parameters. We conjecture this to hold 
for possibly all theories and find confirmation for this conjecture 
in a perturbative analysis. We have computed the running coupling 
from the free energy of a static quark-antiquark pair in the medium 
and find that it exhibits some universal features for large classes of 
models. Finally, we have considered a rotating quark in the plasma. 
Its energy radiation is very similar to the case of $\mathcal{N}\!=\!4$ SYM 
for all models that we study. 

\section*{Acknowledgments}

The work of K.\,S.\ was supported in part by the International Max Planck 
Research School for Precision Tests of Fundamental Symmetries. 
This work was supported by the Alliance Program of the
Helmholtz Association (HA216/EMMI).

\end{document}